\documentclass[twoside]{dis04}
\usepackage{amsmath}
\def\runauthor{M. Hirai, S. Kumano, T.-H. Nagai}
\def\shorttitle{Nuclear parton distribution functions
                and their effects on $\sin^2 \theta_W$ anomaly}
\begin{document}
\title{Nuclear parton distribution functions \\ 
       and their effects on $\sin^2 \theta_W$ anomaly}

\author{M. Hirai$^*$, S. Kumano$^\dagger$, T.-H. Nagai$^\dagger$}
\address{$^*$ Institute of Particle and Nuclear Studies, KEK \\ 
              1-1, Ooho, Tsukuba, 305-0801, Japan \\
         $^\dagger$ Department of Physics, Saga University,
                    Saga, 840-8502, Japan}
\maketitle

\abstracts{Nuclear parton distribution functions (NPDFs) are investigated by
      analyzing the data on structure functions $F_2^A$ and Drell-Yan cross
      sections $\sigma_{DY}^{pA}$. An important point of this analysis is
      to show uncertainties of the NPDFs by the Hessian method.
      The analysis indicates that the uncertainties are large for 
      antiquark distributions at $x>0.2$ and gluon distributions
      in the whole $x$ region. We also discuss a nuclear effect on
      the NuTeV $\sin^2 \theta_W$ anomaly as an application.}
      
\vspace{-7.9cm}
{\hfill SAGA-HE-209-04}
\vspace{+7.0cm}

\section{Introduction}

We report an analysis on nuclear parton distribution functions (NPDFs)
and its application to the NuTeV $\sin^2 \theta_W$ issue. Although
the PDFs in the nucleon are now known rather accurately, their nuclear
modifications are not well determined. It is important to determine
the NPDFs not only for establishing QCD in nuclei but also for applications
to heavy-ion physics and neutrino reactions. 

We proposed the optimum NPDFs in 2001 by a $\chi^2$ analysis of
nuclear data on deep inelastic lepton scattering \cite{saga01}.
In recent years, it becomes important to show reliability of
the obtained PDFs. Actually, uncertainties of the PDFs have been
investigated in unpolarized and polarized PDFs in the nucleon.
Since there was no serious error estimate of the NPDFs, we investigated
the uncertainties by the Hessian method in the recent version 
\cite{npdf04}. In addition, Drell-Yan and HERMES data
are added into the data set, and charm-quark distributions are included
in the new analysis. 

The NuTeV collaboration reported that their neutrino scattering
experiments indicate a significant deviation for the weak mixing angle
$\sin^2 \theta_W$ from other measurements. Because the target is
the iron nucleus, nuclear effects need to be investigated before
discussing any new physics mechanisms. In particular, we study
effects due to the difference between $u_v$ and $d_v$ nuclear 
modifications \cite{sinth02,sinth04}.

This paper consists of the following sections. In Sec. \ref{method},
the $\chi^2$ analysis method is explained for obtaining the optimum NPDFs.
Results on the NPDFs are shown in Sec. \ref{results}, and
their effects are discussed on the determination of $\sin^2 \theta_W$
in Sec. \ref{effects}.

\section{Analysis method}
\label{method}
                
The NPDFs are expressed in terms of parameters, which are determined
by an analysis of the data on nuclear structure functions $F_2^A$ and
Drell-Yan processes. 
The NPDFs are taken as a nucleonic PDF multiplied by a weight function $w_i$
which indicates nuclear modifications:
\begin{equation}
f_i^A (x, Q_0^2) = w_i(x,A,Z) \, f_i (x, Q_0^2),
\label{eqn:paramet}
\end{equation}
where $A$ and $Z$ are mass number and atomic number, and
$i$ denotes a parton-distribution type.
The function $w_i$ is expressed by parameters:
\begin{equation}
w_i(x,A,Z) = 1+\left( 1 - \frac{1}{A^{1/3}} \right) 
          \frac{a_i(A,Z) +b_i x+c_i x^2 +d_i x^3}{(1-x)^{\beta_i}}.
\label{eqn:wi}
\end{equation}
The $A$ dependence is motivated by a simple picture of nuclear volume
and surface contributions to cross sections. The cubic functional
form for the $x$ dependence is motived by the shape of $F_2^A/F_2^D$ data,
and the factor $1/(1-x)^{\beta_i}$ is introduced to reproduce
the Fermi-motion part.

For the NPDFs ($f_i^A$), we take $u_v^A$, $d_v^A$, $\bar q^A$, and $g^A$,
and the initial point is chosen $Q_0^2$=1 GeV$^2$. These distributions
are evolved to experimental $Q^2$ points to calculate 
$\chi^2 = \sum_j (R_{j}^{data}-R_{j}^{theo})^2 / (\sigma_j^{data})^2$,
where $\sigma_j^{data}$ is an experimental error and
$R_j$ indicates the ratios $F_2^A/F_2^{A'}$ and 
$\sigma_{DY}^{pA}/\sigma_{DY}^{pA'}$.
Leading-order expressions are used in the theoretical calculations.
By the $\chi^2$ analysis, we obtain the optimum NPDFs and
a Hessian matrix $H$. Using this matrix,
we calculate the uncertainty of the distribution $f^A(x)$ by 
\begin{equation}
        [\delta f^A(x)]^2 = \Delta \chi^2   \sum_{i,j} 
          \left( \frac{\partial f^A(x,\xi)}{\partial \xi_i} 
                             \right)_{\xi=\hat\xi}
        H_{ij}^{-1}
          \left( \frac{\partial f^A(x,\xi)}{\partial \xi_j} 
                             \right)_{\xi=\hat\xi}
\, ,
        \label{eq:dnpdf}
\end{equation}
where $\xi_i$ is a parameter, and $\hat \xi$ indicates the parameter set
for the minimum $\chi^2$. The $\Delta \chi^2$ determines a confidence region,
and it is taken $\Delta \chi^2$=10.427 for nine parameters \cite{npdf04}.
It corresponds to the one-$\sigma$-error range.

\section{Results on nuclear PDFs}
\label{results}

The experimental data, which are used for the analysis, are taken
for the nuclei: deuteron (D), 
helium-4 ($^4$He), lithium (Li), beryllium (Be), carbon (C),
nitrogen (N), aluminum (Al), calcium (Ca), iron (Fe), copper (Cu),
krypton (Kr), silver (Ag), tin (Sn), xenon (Xe), tungsten (W), gold (Au),
and lead (Pb). The total number of the data is 951.
The minimum $\chi^2$ becomes $\chi^2_{min}$=1489.8 in the current analysis.

\begin{figure}[h!]
\vspace{-0.0cm}
\begin{center}
     \includegraphics[width=0.35\textwidth]{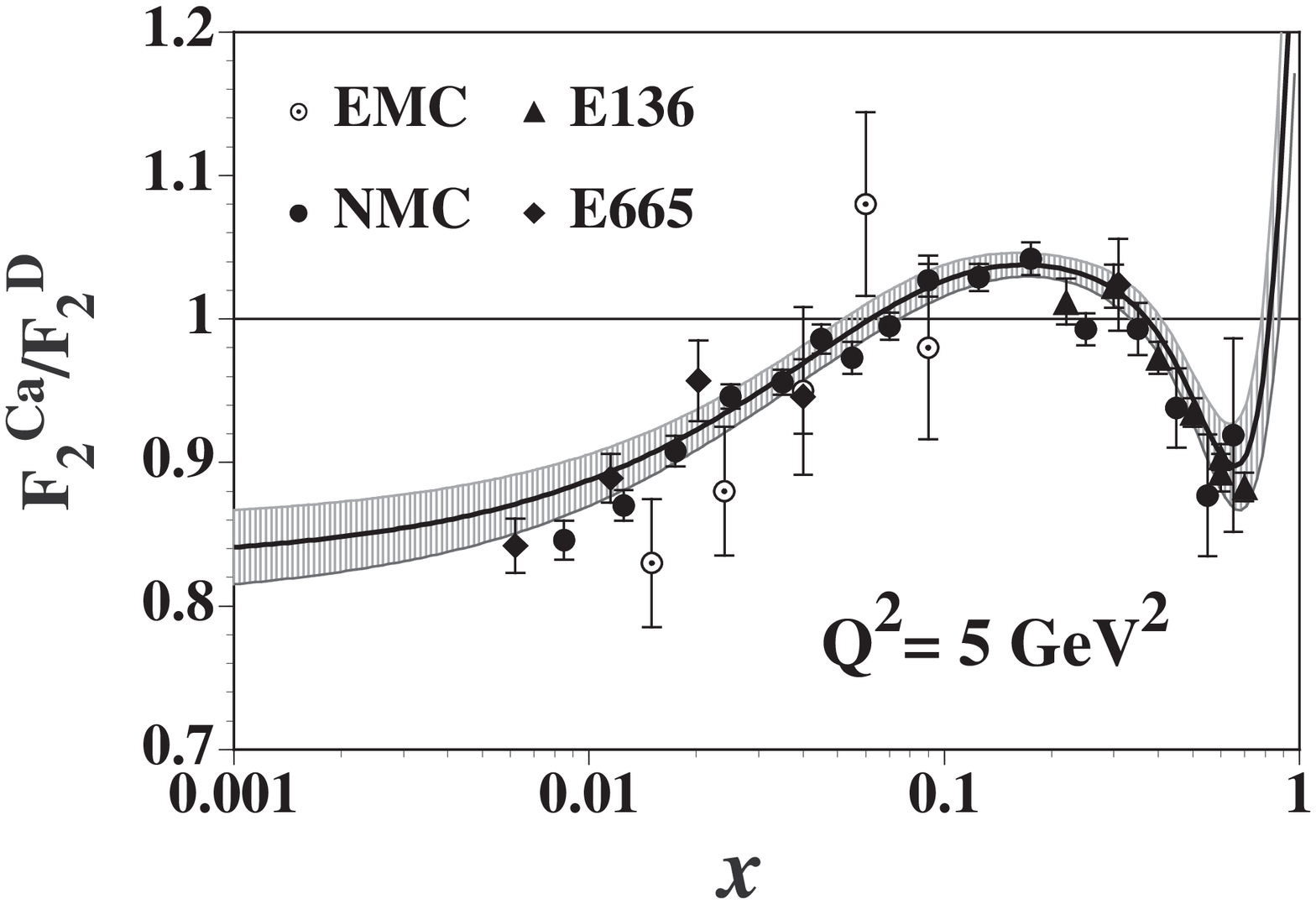}
\hspace{0.3cm}
     \includegraphics[width=0.35\textwidth]{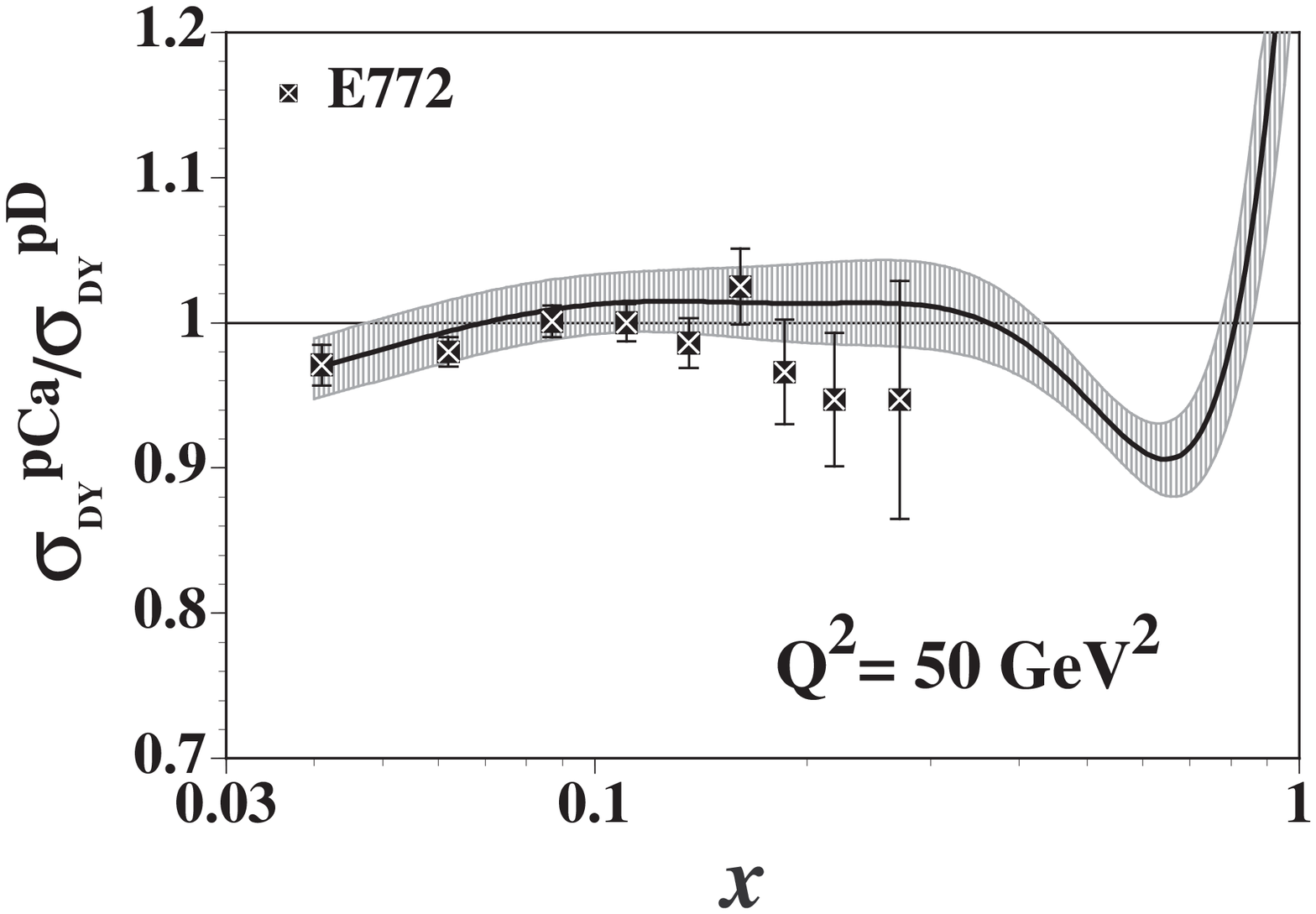}
\end{center}
\vspace{-0.2cm}
\caption{Fit results are compared with the data for $F_2^{Ca}/F_2^D$
         and $\sigma_{DY}^{pCa}/\sigma_{DY}^{pD}$. The curves and
         shaded areas indicate the fit results and their uncertainties
         at $Q^2$=5 or 50 GeV$^2$.}
\vspace{-0.4cm}
\label{fig:f2-dy}
\end{figure}

Among many nuclear data sets, we show only two examples
in Fig. \ref{fig:f2-dy} for comparing fit results with the data
for the calcium nucleus. The parametrization results and uncertainties
are calculated at $Q^2$=5 and 50 GeV$^2$ for $F_2$ and
Drell-Yan, respectively. Because the data are taken at various $Q^2$ points,
the curves cannot be really compared with the data due to
the $Q^2$ difference. However, because the scaling violation is not
a large effect, it is obvious in Fig. \ref{fig:f2-dy} that the analysis well
reproduces the data. The $F_2$ data play a role of determining antiquark
distributions at small $x$ and valence-quark distributions in the medium-
and large-$x$ regions. On the other hand, the Drell-Yan data provide
a constraint on the antiquark distributions at $x \sim 0.1$.

\begin{figure}[h!]
\vspace{-0.2cm}
\begin{center}
     \includegraphics[width=0.32\textwidth]{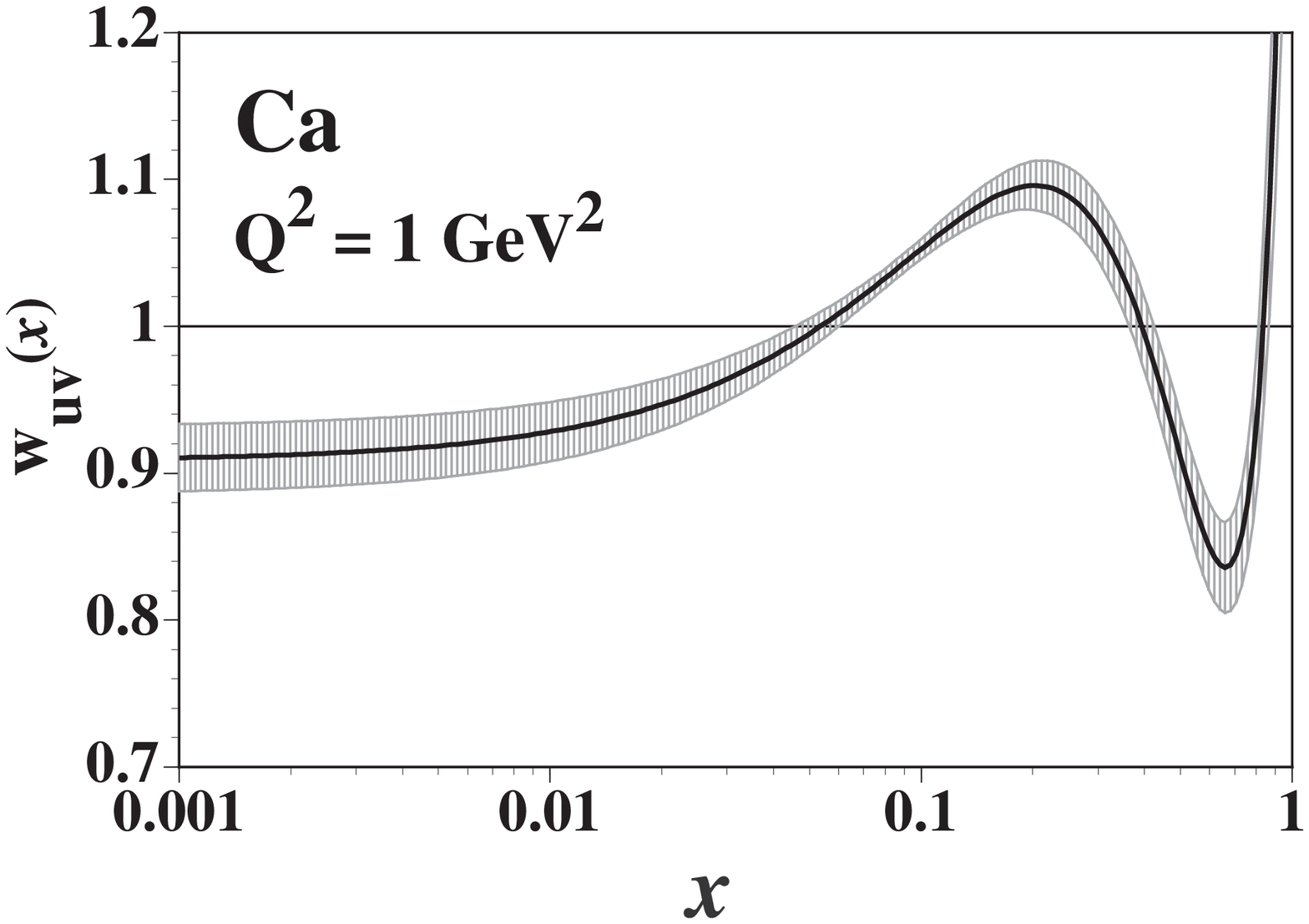}
\hspace{0.0cm}
     \includegraphics[width=0.32\textwidth]{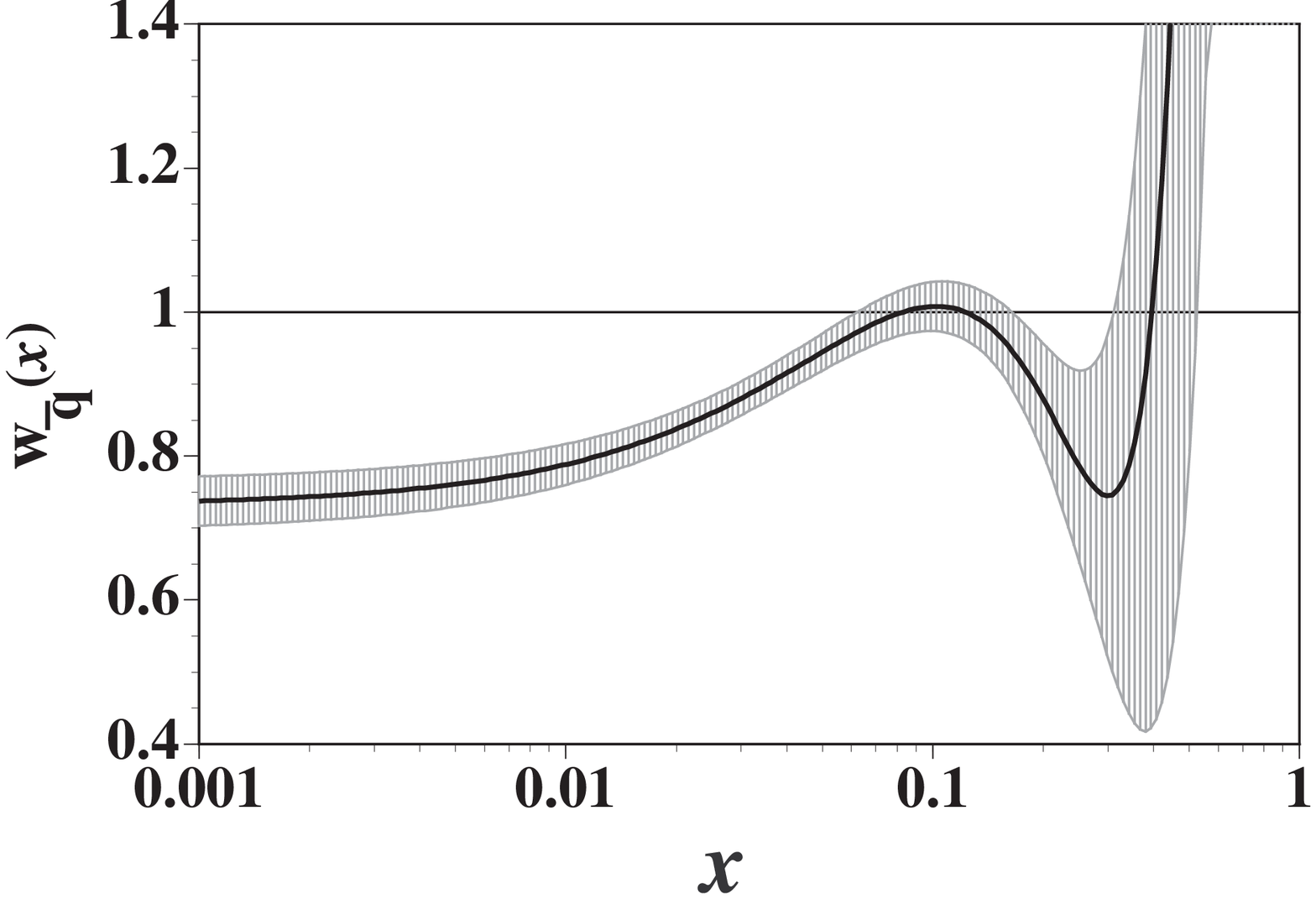}
\hspace{0.0cm}
     \includegraphics[width=0.32\textwidth]{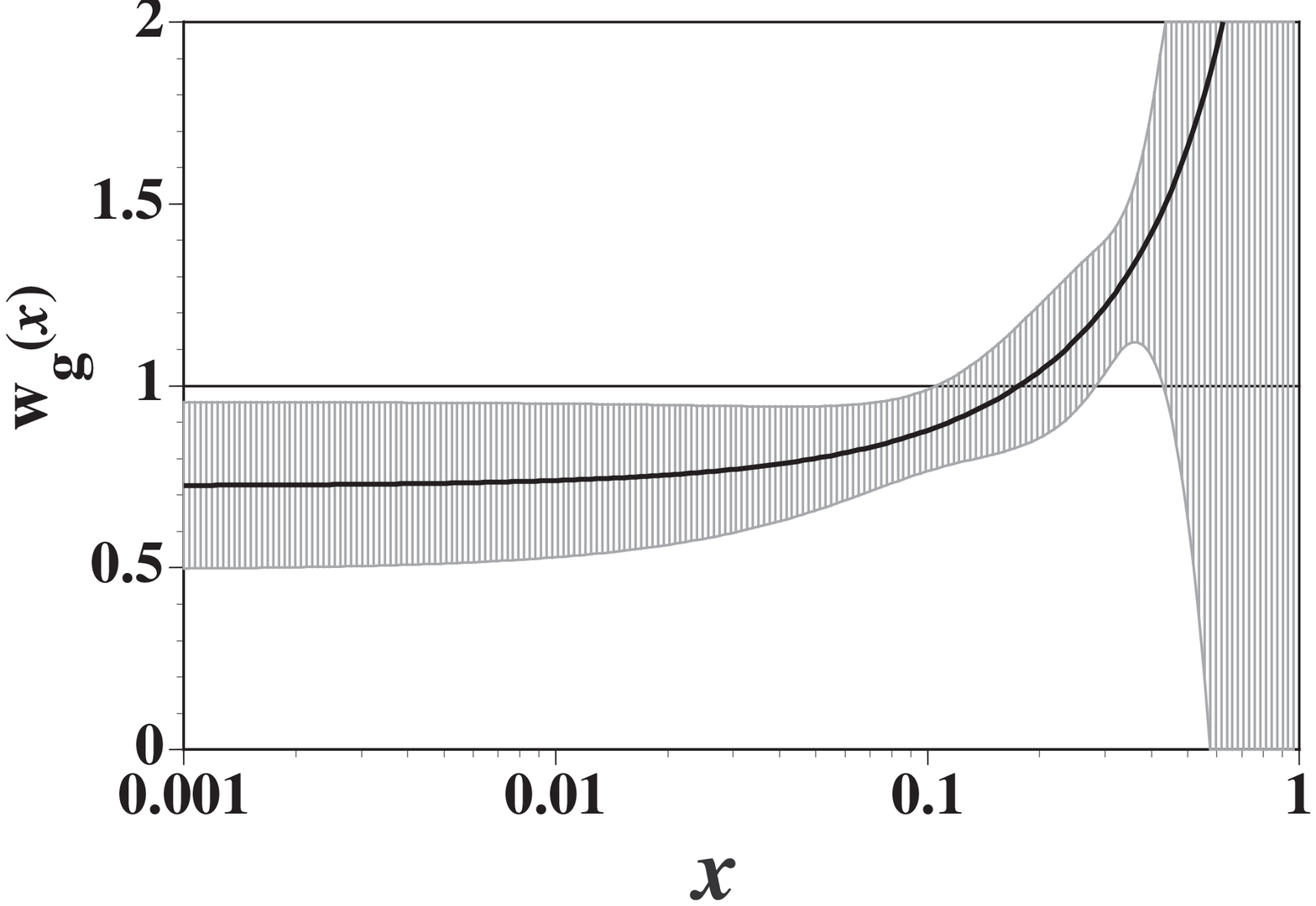}
\end{center}
\vspace{-0.2cm}
\caption{Weight functions $w_{u_v}$, $w_{\bar q}$, and $w_g$
         and their uncertainties are shown for the calcium nucleus
         at $Q^2$=1 GeV$^2$.}
\vspace{-0.3cm}
\label{fig:w-ca}
\end{figure}

From the $\chi^2$ analysis, we obtain the NPDFs for nuclei
with $A=$2 to about 208. Among them, we show nuclear modifications
of a medium size nucleus, calcium. In Fig. \ref{fig:w-ca}, the weight
functions are shown for the distributions $u_v^{Ca}$, $\bar q^{Ca}$,
and $g^{Ca}$ at $Q^2$=1 GeV$^2$. The shaded areas
indicate uncertainties also at $Q^2$=1 GeV$^2$.
The antiquark distributions at small $x$ and 
the valence-quark distributions are well determined. 
However, the antiquark distributions at $x>0.2$ and gluon
distributions in the whole $x$ region have large uncertainties.
The valence-quark distributions are determined by the $F_2$
data at medium $x$, and the small-$x$ part is constrained
by the baryon-number and charge conservations. For fixing $\bar q^A$
at large $x$ and $g^A$, we need new data which are sensitive to
these distributions \cite{dy}.

\vspace{-0.6cm}
\section{Effects on NuTeV $\sin^2 \theta_W$ anomaly}
\label{effects}

The NuTeV collaboration suggested that their neutrino scattering data
should indicate a significant deviation of $\sin^2 \theta_W$ value
from other measurements. Their value is
$\sin^2 \theta_W = 0.2277 \pm 0.0013 \, \text{(stat)} 
                         \pm 0.0009 \, \text{(syst)}$
in comparison with a global analysis of other data: 
$\sin^2 \theta_W= 0.2227 \pm 0.0004$. Because their experiments use
the iron target, nuclear corrections have to be investigated.
Among various nuclear corrections, it was shown in Ref. \cite{sinth02}
that the difference between nuclear modifications of $u_v$ and $d_v$
should contribute to the determination of $\sin^2 \theta_W$. 
We investigate more details by using $\chi^2$ analysis results
for nuclear valence-quark distributions.

For extracting $\sin^2 \theta_W$, the following Paschos-Wolfenstein (PW)
relation is useful:
$ R^-  = ( \sigma_{NC}^{\nu N}  - \sigma_{NC}^{\bar\nu N} )/
         ( \sigma_{CC}^{\nu N}  - \sigma_{CC}^{\bar\nu N} )
        =  1/2 - \sin^2 \theta_W $,
where $\sigma_{CC}^{\nu N}$ and $\sigma_{NC}^{\nu N}$ indicate
deep inelastic neutrino-nucleon cross sections
for charged-current and neutral-current processes, respectively.
This relation is derived for the isoscalar
nucleon without strange-antistrange asymmetry ($s-\bar s=0$). 
We should note that the NuTeV experiments have been done for 
the iron nucleus which has a neutron excess. There could be various
nuclear corrections to the PW relation. In fact, writing down
the relation for a nuclear target, we obtain
\begin{align}
& R_A^- =  \bigg [ \,
\bigg ( \frac{1}{2} - \sin^2 \theta_W \bigg ) \,
\{ 1+ \varepsilon_v (x) \, \varepsilon_n (x) \}
+\frac{1}{3} \, \sin^2 \theta_W \, 
\{ \varepsilon_v (x)  + \varepsilon_n (x) \}
\nonumber \\
       &  \ \ \ \ \ 
+ \bigg ( \frac{1}{2} - \frac{2}{3} \, \sin^2 \theta_W \bigg ) \, 
  \varepsilon_s (x)
+ \bigg ( \frac{1}{2} - \frac{4}{3} \, \sin^2 \theta_W \bigg ) \,
  \varepsilon_c (x) \bigg ] \, 
\bigg /  \, \bigg [ \,
1+ \varepsilon_v (x) \, \varepsilon_n (x)
\nonumber \\
       &  \ \ \ \ \ 
+ \frac{1+(1-y)^2}{1-(1-y)^2} 
 \{ \varepsilon_v (x) + \varepsilon_n (x)  \}
+\frac{2 \{ \varepsilon_s (x) - (1-y)^2 \varepsilon_c (x) \}}{1-(1-y)^2} 
\, \bigg ]
\, .
\label{eqn:apw2}
\end{align}
Here, the correction factors $\varepsilon_v$, $\varepsilon_n$,
$\varepsilon_s$, and $\varepsilon_c$ are defined by 
$\varepsilon_v = (w_{d_v}-w_{u_v})/(w_{d_v}+w_{u_v})$,
$\varepsilon_n = ((N-Z)/A) \, (u_v-d_v)/(u_v+d_v)$,
$\varepsilon_s = s_v^A/[w_v \, (u_v+d_v)]$, and
$\varepsilon_c = c_v^A/[w_v \, (u_v+d_v)]$,
where $N$ is the neutron number and $q_v$ is defined by $q_v=q-\bar q$. 
The functions $w_{u_v}$ and $w_{d_v}$ indicate nuclear modifications for
$u_v$ and $d_v$ distributions as explained in section \ref{method}.
The function $w_v$ is defined by the average, $w_v = (w_{d_v}+w_{u_v})/2$.
We note that Eq. (\ref{eqn:apw2}) becomes the PW relation 
in the limit $\varepsilon \rightarrow 0$.

In particular, we investigate $\varepsilon_v$ effects on
the $\sin^2 \theta_W$ determination \cite{sinth02,sinth04}.
For calculating $\varepsilon_v$ effects, we should take the NuTeV kinematical
into account. Such a kinematical effect can be incorporated by using
functionals provided by the NuTeV collaboration. 
The nuclear modification ($\varepsilon_v$) effects on the $\sin^2 \theta_W$
are then calculated. Our research is still in progress \cite{sinth04}; 
however, the preliminary result indicates a large uncertainty for
the $\sin^2 \theta_W$ determination. It means that the nuclear modification
could not be determined accurately and that there is still a
possibility that the deviation is explained by the nuclear effect.

\vspace{-0.05cm}
\section{Summary}
\label{summary}
\vspace{-0.2cm}

Nuclear parton distribution functions (NPDFs) have been obtained by analyzing
the data for the structure function $F_2$ and Drell-Yan cross sections.
Their uncertainties were calculated by the Hessian method. The uncertainties
are especially large for the antiquark distributions at $x>0.2$ and
the gluon distributions, so that they need future experimental measurements.
Such a determination of the NPDFs could contribute to the clarification
of the NuTeV $\sin^2 \theta_W$ anomaly.

\vspace{-0.05cm}
\section*{Acknowledgements} 
\vspace{-0.2cm}
S.K. was supported by the Grant-in-Aid for Scientific Research from
the Japanese Ministry of Education, Culture, Sports, Science, and Technology. 

\vspace{-0.05cm}


\end{document}